\documentclass[final,3p]{elsarticle}

\usepackage{multirow}
\usepackage{siunitx}
\usepackage{tikz}
\usetikzlibrary{decorations.markings, arrows}

\usepackage{graphicx,times}
\usepackage{latexsym}
\usepackage{mathtools}

\usepackage{amsmath,amssymb,amsbsy,amsfonts}
\usepackage{array}
\usepackage{bm}
\usepackage{graphics}
\usepackage{mathrsfs}
\usepackage{xcolor}
\usepackage{cancel}
\usepackage[normalem]{ulem}

\usepackage{hyperref}
\usepackage[capitalise]{cleveref}

\journal{Physics Letters B}

\begin{document}

\begin{frontmatter}



\title{Emergence of Gauss' Law in a $Z_2$ Lattice Gauge Theory in $1+1$ Dimensions}
\author[SC]{Jernej Frank}
\author[EH]{Emilie Huffman}
\author[SC]{Shailesh Chandrasekharan}
\address[SC]{Department of Physics, Box 90305, Duke University,
  Durham, NC 27708, USA}
\address[EH]{University of Wurzburg, Germany}

\begin{abstract}
  We explore a $Z_2$ Hamiltonian lattice gauge theory in one spatial dimension with a coupling $h$, without imposing any Gauss' law constraint. We show that in our model $h=0$ is a free deconfined quantum critical point containing massless fermions where all Gauss' law sectors are equivalent. The coupling $h$ is a relevant perturbation of this critical point and fermions become massive due to confinement and chiral symmetry breaking. To study the emergent Gauss' law sectors at low temperatures in this massive phase we use a quantum Monte Carlo method that samples configurations of the partition function written in a basis in which local conserved charges are diagonal. We find that two Gauss' law sectors, related by particle-hole symmetry, emerge naturally. When the system is doped with an extra particle, many more Gauss's law sectors related by translation invariance emerge. Using results in the range $0.01 < h \leq 0.15 $ we find that three different mass scales of the model behave like $h^{p}$ where $p \approx 0.579$.
\end{abstract}



%%\PACS 71.10.Fd,02.70.Ss,11.30.Rd,05.30.Rt

%\end{keyword}

\end{frontmatter}



\section{Introduction}
\label{sec1}

The possibility of using quantum computers to understand quantum field theories has become an exciting field of research lately \cite{Zohar:2016dfx}. One of the challenges for applications to nuclear and particle physics is our ability to formulate all local quantum fields on a finite dimensional Hilbert space \cite{Wiese:2013uua}. This is commonly referred to as "field digitization". However, from the perspective of renormalization group flows it is not clear if a field digitization onto a finite number of qubits will always preserve the properties of the continuum quantum field theory we wish to study \cite{Alexandru:2019nsa}. When it indeed does we can view it as a new way to regulate the continuum theory. With this perspective, field digitization was also recently given the name "qubit-regularization" \cite{PhysRevD.100.054505}. While there are many ways to accomplish this for gauge theories \cite{Chandrasekharan:1996ih,PhysRevD.91.054506}, the low energy physics that emerges can depend on the details of the Hilbert space formulation \cite{PhysRevB.97.205108}. Simple quantum field theories are currently being formulated so that they can be studied using a quantum computer \cite{Zohar:2015hwa,Pichler:2015yqa,Martinez:2016yna,PhysRevA.98.032331,Evans:2018njs,Kaplan:2018vnj,Unmuth-Yockey:2018xak,PhysRevA.100.012320,Alexandru:2019ozf}. One of the main long-term challenges is to be able to formulate and study strongly coupled gauge theories like QCD \cite{Brower:2003vy}. One of the goals of our work is to understand the physics of a simple qubit-regularized quantum field theory with some similarities to QCD, but simple enough to be simulated on a quantum computer in the near future. 

Since the Hamiltonian of a gauge theory is invariant under local symmetry transformations, the Hilbert space of states can be divided into sectors labeled by local conserved gauge charges \cite{PhysRevD.11.395}. Under time evolution these sectors do not mix with each other and each sector satisfies its own Gauss' law. For example, if the local conserved gauge charge density is chosen to be $\rho_G(\bf r)$ in quantum Electrodynamics, the Gauss' law will take the form ${\pmb{\nabla}} \cdot {\bf E}({\bf r},t) \ -\  \rho({\bf r},t)\ =\ \rho_G({\bf r})$. Using Maxwell's equations we learn that the sector of the quantum Hilbert space with $\rho_G({\bf r}) = 0$ is the physical one. Imposing this local constraint on the allowed space of states is an important step in the formulation of the quantum gauge theory.

In this work we explore a very simple lattice gauge theory; much simpler even than the Schwinger model, that is often studied in the context of quantum computation. The goal of our work is to understand the physics of this simpler model if we do not even impose the Gauss' law constraint, since imposing it can be difficult when formulating a gauge theory on a quantum computer \cite{PhysRevA.98.032331}. Furthermore, we explore if such constraints can emerge naturally at low temperatures without imposing them. Our model is a simple one dimensional $Z_2$ lattice gauge theory which contains massless fermions interacting with a $Z_2$ lattice gauge fields. Such gauge theories in higher dimensions are interesting in condensed matter physics and are believed to describe frustrated quantum magnets and spin liquid phases of materials \cite{Sachdev2008,Balents2010} and have also been studied by other groups recently \cite{PhysRevX.6.041049,Gazit:2018vsa,PhysRevD.98.074503,PhysRevB.96.205104}. The Hamiltonian of our model is given by
\begin{align}
H \ =\  -\ \sum_j \Big\{ (c^\dagger_j c_{j+1} + c^\dagger_{j+1} c_j)\sigma_j^x\ + h \sigma_j^z \Big\}.
\label{eq:model}
\end{align}
Here $c^\dagger_j$ and $c_j$ create and annihilate fermions on the sites $j=0,1..,L-1$ of a periodic lattice and $\sigma^x_j, \sigma_j^z$ are the Pauli matrices that represent the $Z_2$ gauge fields associated to links connecting the sites $j$ and $j+1$. We assume $L$ to be even to preserve particle hole symmetry.

The gauge invariance of our Hamiltonian can be seen from the relation
\begin{align}
[H,Q_j] \ =\ 0,    
\end{align}
where $Q_j = \sigma_{j-1}^z\sigma^z_j(-1)^{n_j}$ are the local charge operators. Here $n_j = c^\dagger_j c_j$. The set of their simultaneous eigenvalues $\{Q_j=\pm 1\}$ labels one of $2^L$ possible Gauss' law sectors. We label each sector with a unique number from $0$ to $2^{L}-1$ using the relation
\begin{align}
Q = \sum_{j=0}^{L-1} (1+Q_j) 2^{j-1},
\label{eq:GLQ}
\end{align} 
and compute the probability distribution $P(Q)$ as a function of temperature. The emergent sectors $Q_e$ are those which have a non-zero probability $P(Q_e)$ at zero temperature. In the next section we will argue that when $h=0$ our lattice model describes a deconfined quantum critical point where all Gauss' law sectors are equivalent. In later sections, using Monte Carlo calculations, we will show that when $h \neq 0$ two Gauss' law sectors related by particle-hole symmetry emerge, fermions are confined and acquire a mass due to chiral symmetry breaking.
These attributes of our model are similar to the Schwinger model, which is often used as a simple toy model of QCD.

The emergence of Gauss' law sectors when $h\neq 0$ was discussed recently and studied using auxiliary field Monte Carlo (AFMC) methods in two spatial dimensions \cite{PhysRevX.6.041049,Gazit:2018vsa}. Unfortunately, it is not possible to compute $P(Q)$ using AFQMC calculations due to sign problems. Since our studies are restricted to one spatial dimension, we can work in the basis in which fermions are represented by their occupation numbers and the $Z_2$ electric field operators ($\sigma_j^z$) are diagonal, without encountering sign problems. Since every configuration is naturally in a fixed Gauss' law sector with a well defined set of charges $\{Q_j\}$, we can easily compute the  probability distribution $P(Q)$ using our Monte Carlo method. We can also address the effects of doping the system away from the particle-hole symmetric situation. We show that $L$ different Gauss' law sectors related by translation symmetry emerge if we dope the system with one extra particle.

\begin{figure}
  \begin{minipage}[t]{0.45\textwidth}
\begin{center}
\includegraphics[width=\textwidth]{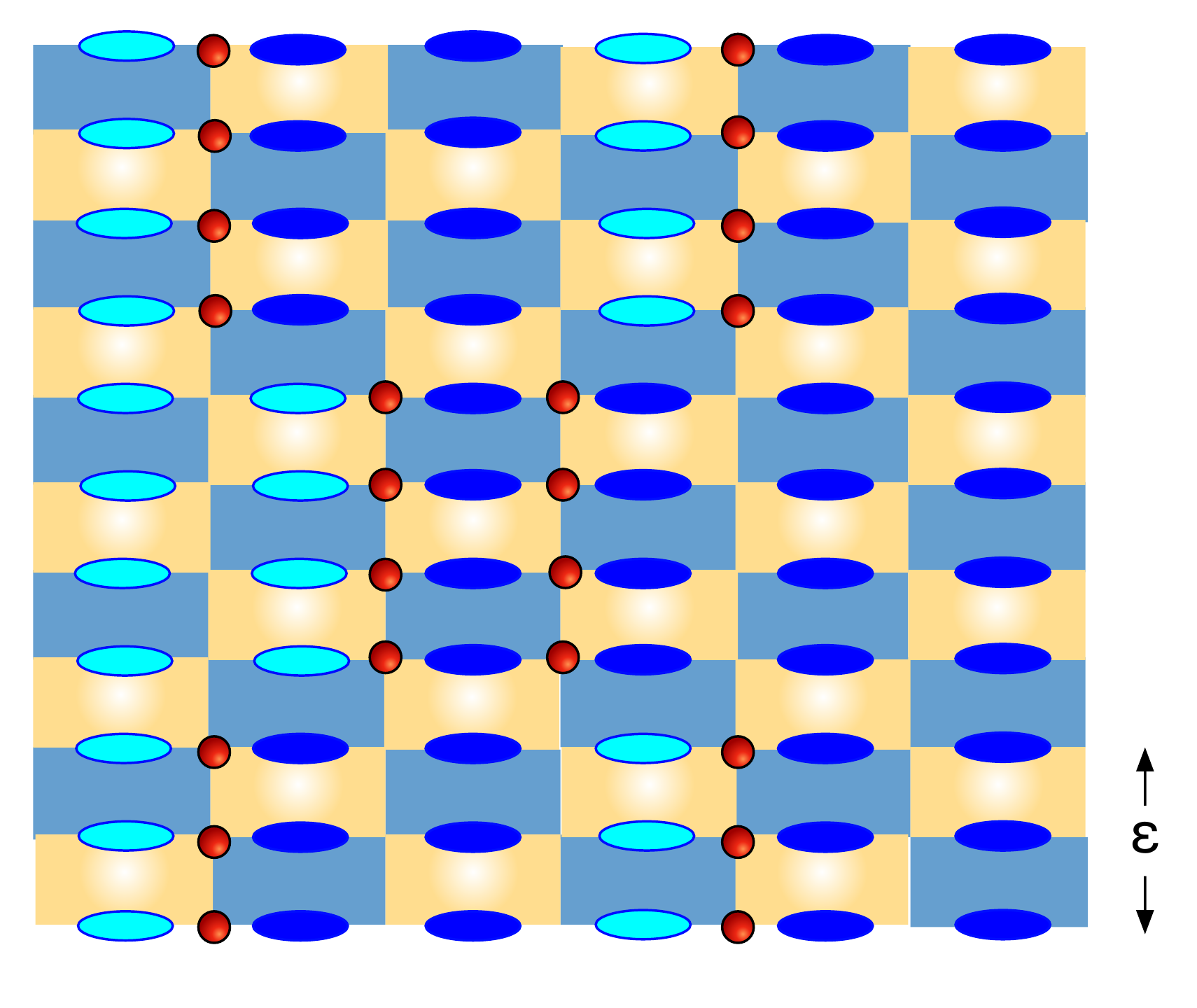}
\caption{\label{fig:fig1} Illustration of a worldline configuration of fermions (circles) and gauge fields (dark ($\sigma^z_j=+1$) and light ($\sigma^z_j=-1$) ovals). Transfer matrix elements $\mathrm{e}^{\varepsilon H}$ are divided into fermion hopping (dark plaquettes) and the coupling $h$ (light plaquettes). Our time discretization error is chosen to be  $\varepsilon=0.1$.}
\end{center}
\end{minipage}
\hspace{1.5cm}
\begin{minipage}[t]{0.45\textwidth}
\begin{center}
\includegraphics[width=\textwidth]{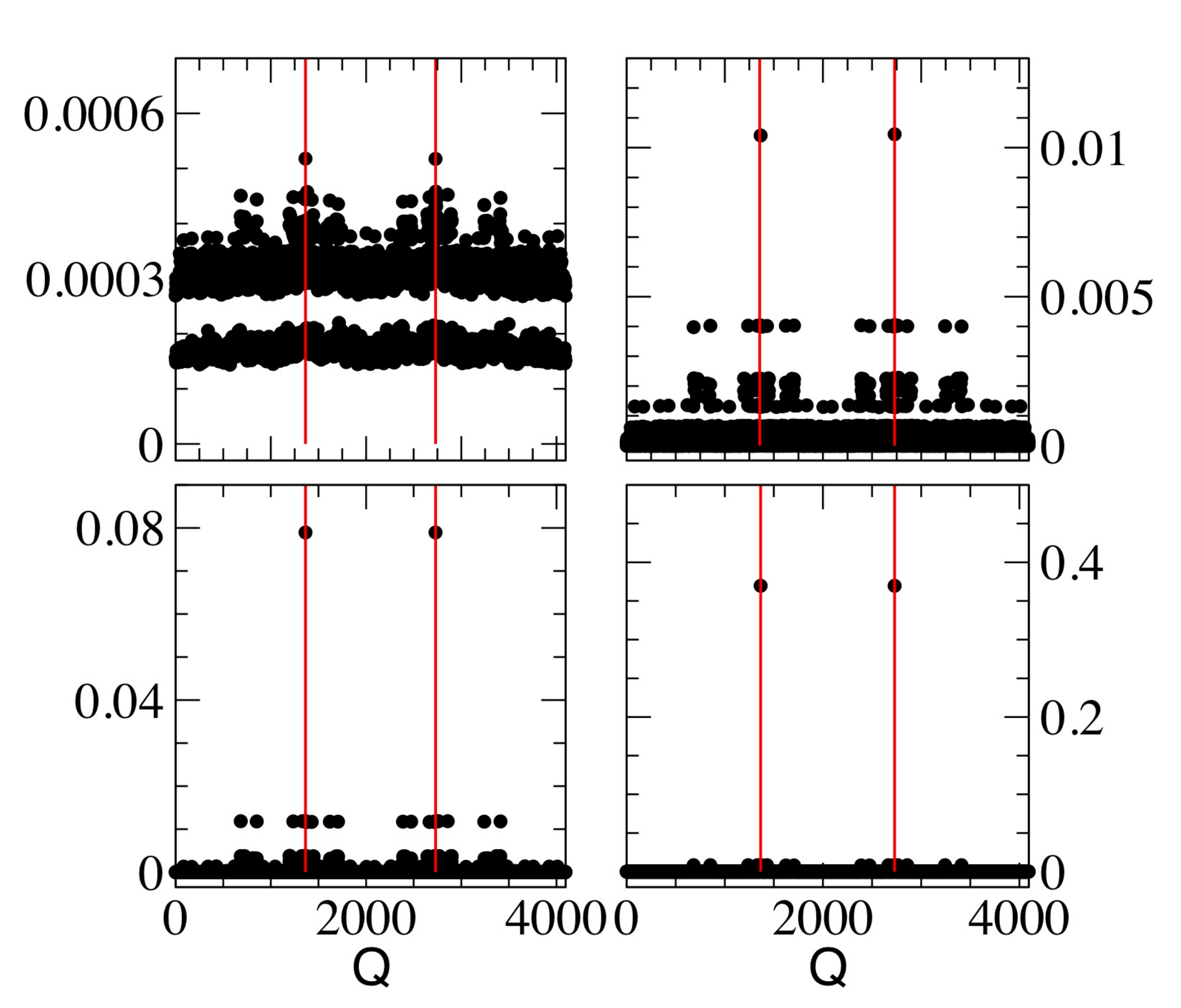}
\caption{\label{fig:fig2} Plot of $P(Q)$ as a function of $Q$ when $h=0.05$ and $L=12$. The four graphs shown are when $\beta=5$ (top left), $\beta=25$(top right), $\beta=50$ (bottom left) and $\beta=100$ (bottom right). The emergent sectors $Q_e=1365, 2730$ are shown as solid vertical lines in the plots.}
\end{center}
\end{minipage}
\end{figure}

Our calculations are performed in a discrete time formulation of the path integral, where we divide the inverse temperature $\beta$ into equal slices of width $\varepsilon = 0.1$ \cite{Wiese:1992np}. An illustration of the worldline configuration is shown in \cref{fig:fig1}. While the algorithm is straight forward \cite{Syljuaen:2002zz,Prokofev:2001ddj,Chandrasekharan:2008gp}, more details about how we overcome some issues that arise in a gauge theory can be found in the supplementary material.

\section{Deconfined Quantum Critical Point}
\label{sec2}

In order to understand our model \cref{eq:model} better, let us first set $h=0$ and ignore the gauge fields in the fermion hopping term. If we then add a staggered fermion mass term  $H^{(m)}_j = (-1)^{j + n_j}$ to the model it is easy to verify that the resulting free fermion Hamiltonian,
\begin{align}
  H_m \ =\  \sum_j \Big\{ -(c^\dagger_j c_{j+1} + c^\dagger_{j+1} c_j)\ + m H^{(m)}_j \Big\},
\end{align}
describes relativistic staggered fermions with mass $m$ \cite{PhysRevD.16.3031}. In this sense our model describes the physics of massless staggered fermions coupled to $Z_2$ lattice gauge fields.

In fact we can solve our model exactly when $h=0$. To see this, let us define a new set of fermion annihilation operators through the relations $f_0 = c_0$, $f_j = \sigma^x_0\sigma^x_1...\sigma^x_{j-1} c_j$, for $j=1,2,..,L-1$. The new fermion creation operators will naturally be the Hermitian conjugates and $n_j= f^\dagger_j f_j = c^\dagger_j c_j$. It is easy to verify that these new set of fermion operators non only satisfy the usual anti-commutation relations but also commute with the local gauge charges $Q_j = \sigma_{j-1}^z \sigma^z_j (-1)^{n_j}$ defined previously. However, it is important to remember that $Q_j$'s satisfy the constraint,
\begin{align}
Q_0 Q_1...Q_{L-1} = (-1)^{N_f},
\label{eq:Qconst}
\end{align}
where $N_f$ is the total fermion number. This shows that a choice of the Gauss' law sector $\{Q_j \pm 1\}$, also constrains the fermionic Hilbert space.

Let us also define two new operators in the gauge field sector: the Wilson loop operator $W_{L-1} = \sigma_0^x \sigma_1^x... \sigma_{L-1}^x$, and its conjugate $E_{L-1} = \sigma^z_{L-1}$. Although $f_j, f^\dagger_j$ depend on gauge fields they still commute with $W_{L-1}$ and $E_{L-1}$. Note that we can write $\sigma_j^z = E_{L-1} Q_0 Q_1 ...Q_j F_j$ where $F_j = (-1)^{\sum_{k=0}^j n_k}$ is a non-local fermion operator. Hence, we can rewrite \cref{eq:model} as
\begin{align}
  H \  =\  \sum_j \Big\{ -(f^\dagger_j f_{j+1} + f^\dagger_{j+1} f_j) W_j  \ - \
  h E_{L-1} \ Q_0 \ Q_1...\ Q_j \ F_j \Big\},
\label{eq:model1}
\end{align}
where for convenience we define $W_j = 1$ for $j \neq L-1$. Since all fermion operators commute with all $W_j$, $Q_j$ and $E_{L-1}$, this form of the Hamiltonian makes it easy to analyze the physics in each Gauss' law sector. We can work in a basis where all $\{Q_j\}$ are diagonal and the constraint \cref{eq:Qconst} is satisfied. When $h=0$ we observe that the fermions are almost free and massless, except for the coupling to $W_{L-1}$. In a basis where the Wilson loop is diagonal, the choice of $W_{L-1}=\pm 1$ only effects the boundary condition of the free fermion problem and disappears in the thermodynamic limit. Thus, we find that all Gauss' law sectors are equivalent and describe free massless staggered fermions. This implies that $h=0$ is a deconfined quantum critical point in every Gauss' law sector.

\section{Emergence of Gauss' Law}
\label{sec3}

When $h \neq 0$ different Gauss' law sectors have different energies and as far as we know our model cannot be solved exactly. In our work we use Monte Carlo methods to explore the physics. We first study the emergence of Gauss' law at low temperatures when $L=12$ at $h=0.05$ by computing the probability distribution $P(Q)$ for $Q=0,1,...,4095$ at various inverse temperatures $\beta$. In \cref{fig:fig2} we plot $P(Q)$ as a function $Q$ at $\beta=5,25,50,100$. As the temperature reduces, we observe two Gauss' law sectors emerge with $Q_e=1365$ (i.e., $\{Q_j = (-1)^j\}$) and its particle-hole symmetric partner $Q_e=2730$ (i.e., $\{Q_j = (-1)^{j+1}\}$). In these emergent sectors the fermion number is found to satisfy the constraint \cref{eq:Qconst} and is consistent with particle-hole symmetry i.e., $N_f=L/2=6$. These half-filled pattern, $\{Q_j = (-1)^j\}$ and $\{Q_j = (-1)^{j+1}\}$ with $N_f=L/2$ continue to be the emergent sectors even at larger lattice sizes, as predicted in the previous work \cite{PhysRevX.6.041049,Gazit:2018vsa}.

\begin{figure}
  \begin{minipage}[t]{0.45\textwidth}
  \begin{center}
\includegraphics[width=0.98\textwidth]{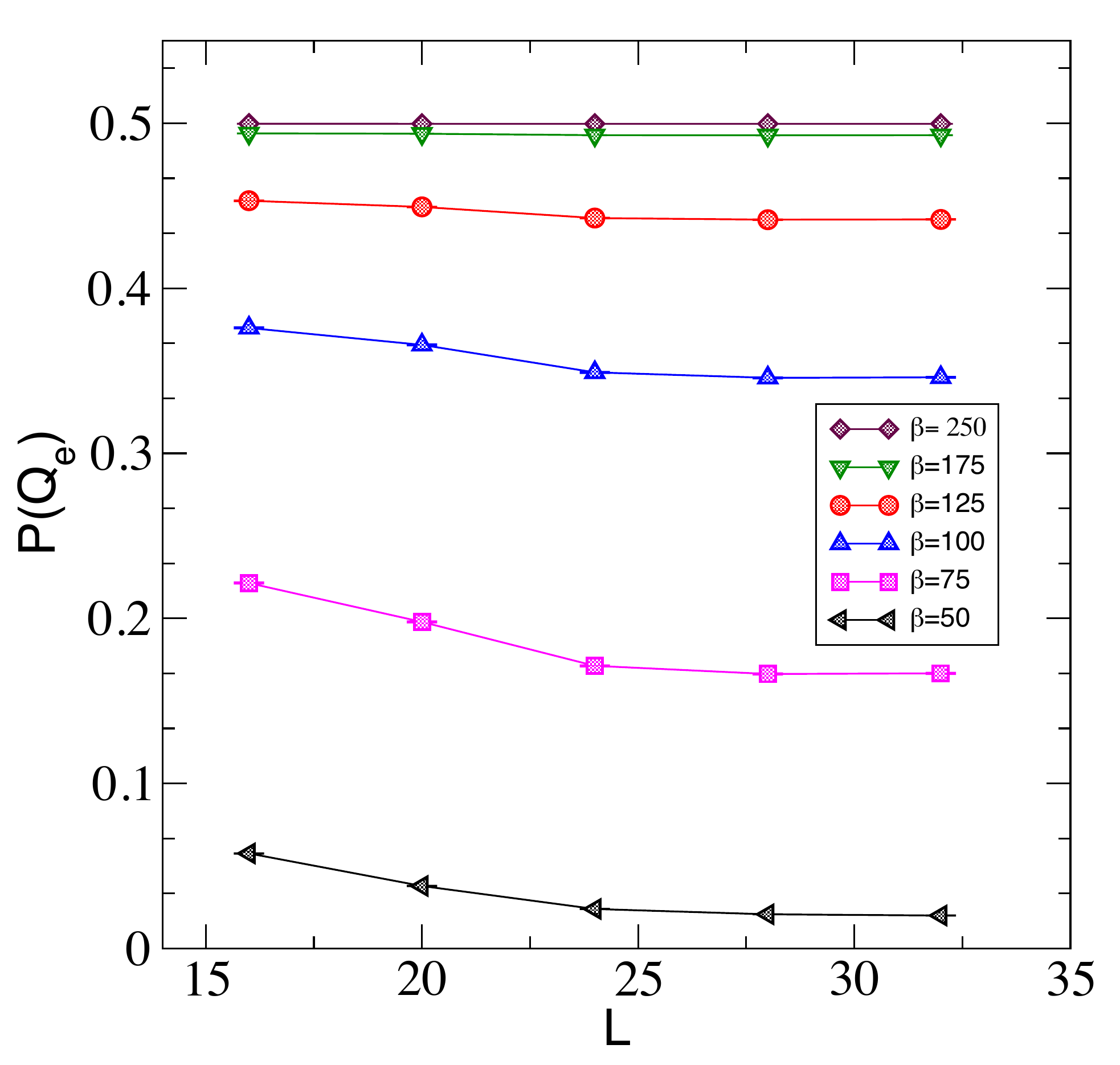}
\caption{\label{fig:fig3} Plot of $P(Q_e)$ as a function of $L$ at various values of $\beta$ when $h=0.05$. Although the number of Gauss' law sectors increase exponentially as $2^L$, $P(Q_e)$ reaches the thermodynamic value for $L > 24$ at the temperatures shown.}
\end{center}
\end{minipage}
\hspace{1.5cm}
\begin{minipage}[t]{0.45\textwidth}
  \begin{center}
  \includegraphics[width=\textwidth]{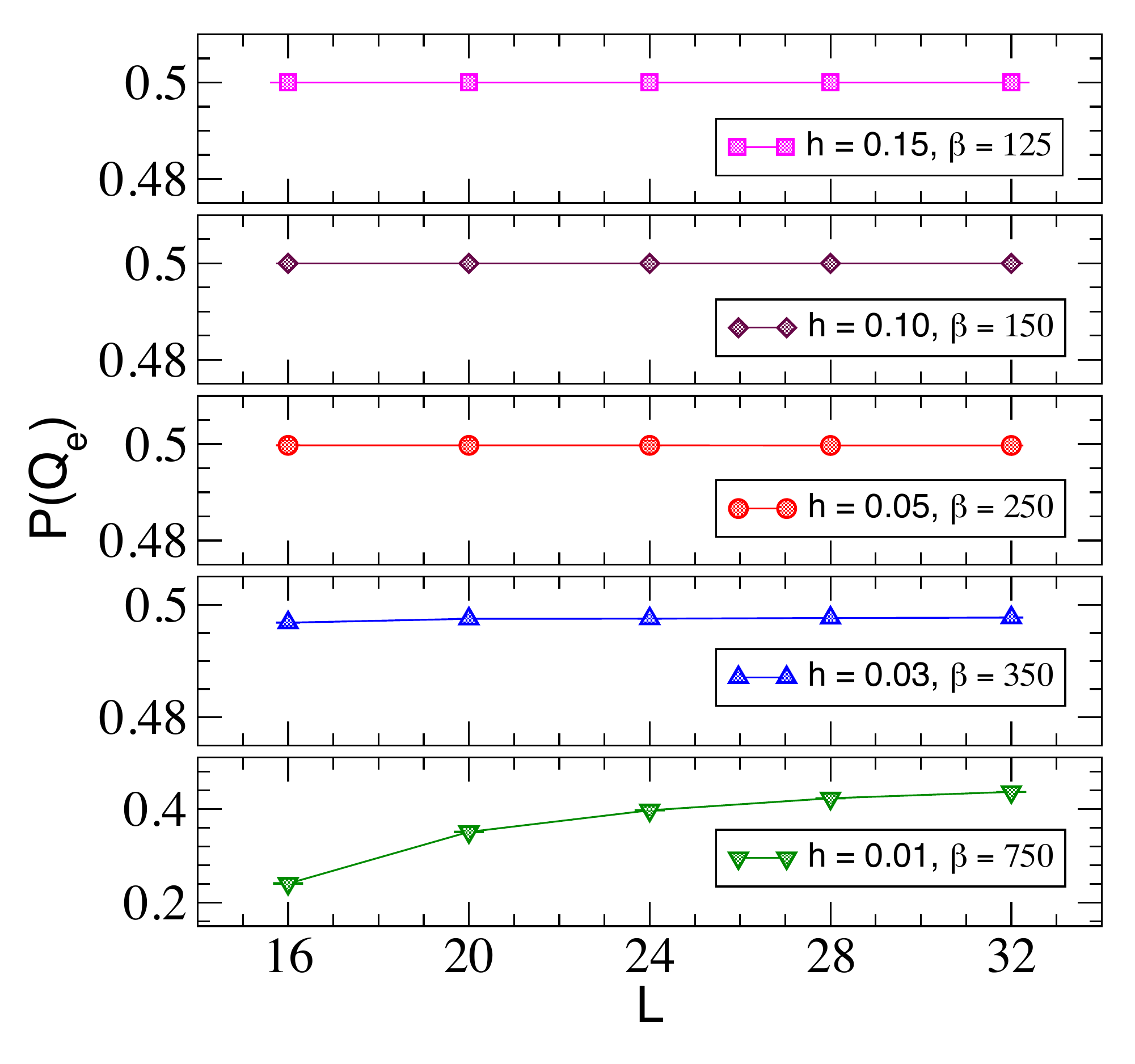}
  \caption{\label{fig:fig4} Plot of $P(Q_e)$ as a function of $L$ at various combinations $(h,\beta)$. Note that $P(Q_e) \approx 0.5$, except when $h=0.01$ where even with $\beta=750$, our smallest temperature, we do yet not see saturation.}
  \end{center}
\end{minipage}
\end{figure}

We then understand how $P(Q_e)$ depends on the temperature and what is the role of the lattice size. In \cref{fig:fig3} we plot the $P(Q_e)$ for each of the two emergent sectors as a function of $L$ for different values of $\beta$ at $h=0.05$. Remarkably, although the number of Gauss' law sectors increase exponentially as $2^{L}$, $P(Q_e)$ saturates to the thermodynamic value for $L > 24$. This implies that in the thermodynamic limit only a few sectors are important at low temperatures. On the other hand very low temperatures are required to project out all the subdominant sectors completely. For example at $h=0.05$ we need $\beta=250$ for $P(Q_e) \approx 0.5$. It is not surprising that this temperature is dependent on $h$ since at $h=0$ all sectors are equally important. In \cref{fig:fig4} we plot $P(Q_e)$ as a function of $L$ at various combinations of $(h,\beta)$. Note that at $h=0.01$, even $\beta=750$ is still not sufficient to project out all the sub-dominant Gauss' law sectors.

Changing $N_f$ must change $Q_e$ as expected from \cref{eq:Qconst}. To study this, we add a chemical potential term $\mu$ to the Hamiltonian and increase it to dope the system with one additional fermion. At $h=0.05$, $L=12$ and $\beta=100$ we observe that when $\mu$ increases from $0$ to $0.2$ the fermion number increases from $N_f=6$ to $N_f=7$ (for a plot of the $\mu$ dependence we refer the reader to the supplementary material). When $N_f=7$ (at $\mu=0.2$)  we observe that twelve Gauss' law sectors given by $Q_e = 3434$, $2773$, $1451$, $2902$, $1709$, $3418$, $2741$, $1387$, $2774$, $1453$, $2906$, $1717$ emerge at low temperatures. In \cref{fig:fig5} we show $P(Q)$ close to these emergent sectors. These sectors are consistent with translation symmetry and satisfy the constraint \cref{eq:Qconst}. The presence of one extra fermion creates two defects in the half filled Gauss' law pattern, that are maximally separated (a pictorial representation of the emergent sectors can be found in the supplementary material).

\section{Chiral Symmetry Breaking}

In \cref{sec2} we argued that the local operator $H^{(m)}_j$ is a fermion mass term in our model. Fortunately, our 
model contains a chiral symmetry that prevents this term. For example it is easy to verify that $H$ is invariant under particle-hole transformations ($\mathbb{C}$) and  translations by one lattice unit ($\mathbb{U}$):
%-----------------------original version: --------------------------
%\begin{align}
%C: \ \ c_j \rightarrow (-1)^j\ c^\dagger_j, \ 
%c^\dagger_j \rightarrow (-1)^j\ c_j,\qquad \qquad
%T: \ \ c_j \rightarrow c_{j+1},\ 
%c^\dagger_j \rightarrow  c_{j+1}.
%\end{align}
%--------------------------------------------------------------------
%-----------------------suggestion:----------------------------------
\begin{align}
\mathbb{C}: \begin{matrix}
    c_j \rightarrow (-1)^j\ c^\dagger_j\\ 
    c^\dagger_j \rightarrow (-1)^j\ c_j
    \end{matrix}\qquad , \qquad
\mathbb{U}: \begin{matrix}
    c_j \rightarrow c_{j+1}\\
    c^\dagger_j \rightarrow  c_{j+1}
    \end{matrix}.
\end{align}
%---------------------------------------------------------------------
It is easy to verify that $H^{m}_j$ is odd under both $\mathbb{C}$ and $\mathbb{U}$. Hence we will refer to both of them together as the chiral symmetry in our model, since preserving at least one of them is necessary to keep the fermions massless. Thus, the ground state expectation value $\phi=\langle H^{(m)}_j\rangle$ can be viewed as the chiral order parameter for this chiral symmetry, when $\phi \neq 0$ both $\mathbb{C}$ and $\mathbb{U}$ must be broken but not necessarily their product. Thus the symmetry that prevents a fermion mass term is a $Z_2$ chiral symmetry in our model.

While the Hamiltonian is invariant under this $Z_2$ chiral symmetry for all values of $h$, it is easy to verify that $\{Q_j = (-1)^j\} \leftrightarrow \{Q_j = (-1)^{j+1}\}$ under both $\mathbb{C}$ and $\mathbb{U}$. Thus, the chiral symmetry is explicitly broken in each of the fixed Gauss law sectors that emerge at low temperatures when $h \neq 0$. This implies that fermions will become massive when $h \neq 0$. There is an additional symmetry related to the transformation $\sigma^z \rightarrow -\sigma^z$ on alternate bonds when $h=0$. This extra symmetry can be used to redefine chiral symmetry in each fixed Gauss' law sector, which helps to keep fermions massless at the deconfined quantum critical point, but not away from it. 

\begin{figure}
     \begin{minipage}[t]{0.45\textwidth}
  \begin{center}
\includegraphics[width=\textwidth]{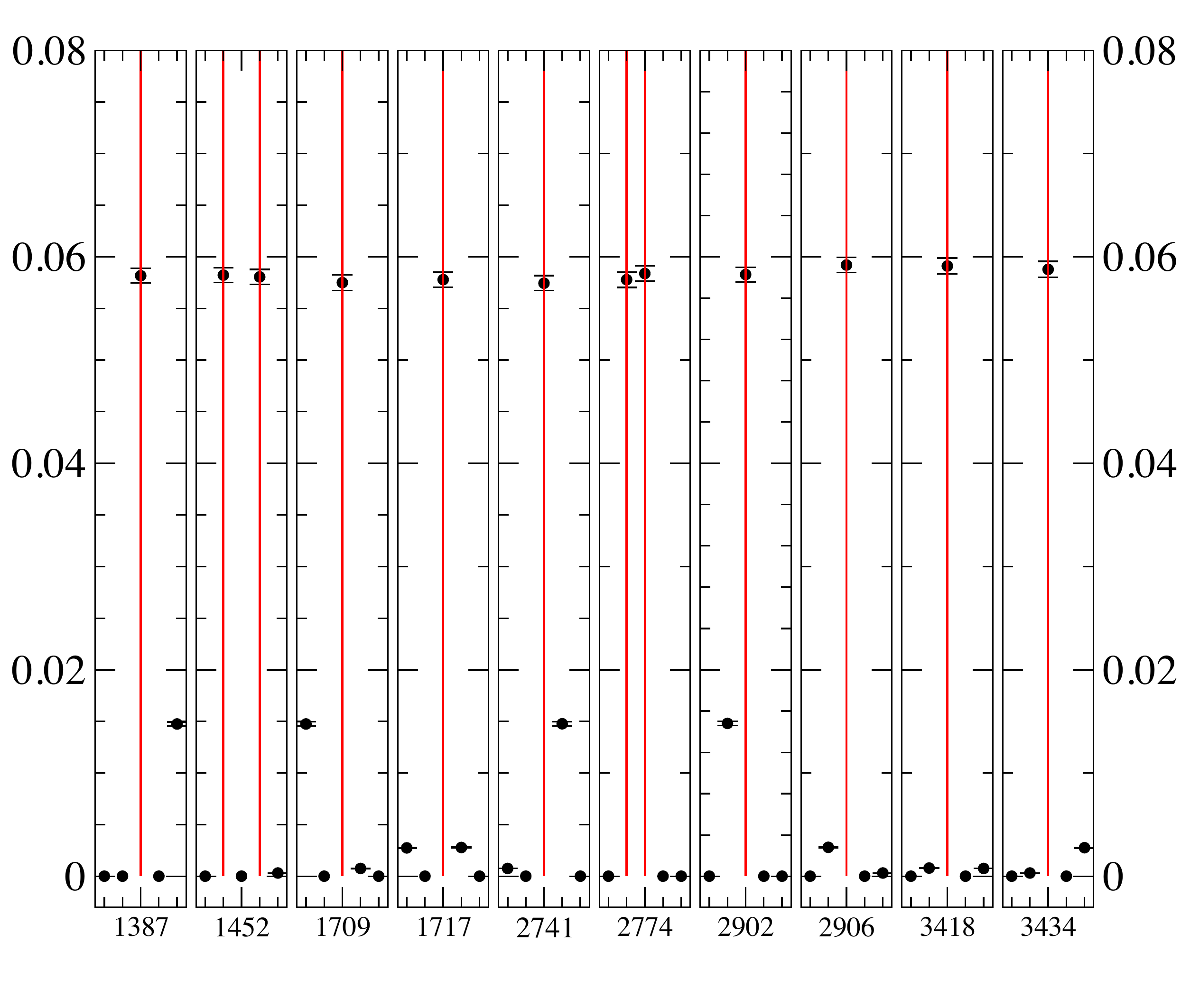}
\caption{\label{fig:fig5} Plot of $P(Q)$ as a function of $Q$ when $h=0.05,\beta=100$ and $L=12$ as in \cref{fig:fig2}, but now without particle-hole symmetry since $N_f=7$.Only values close to $Q_e$ are shown.}
\end{center}
\end{minipage}
\hspace{1cm}
\begin{minipage}[t]{0.45\textwidth}
  \begin{center}
    \includegraphics[width=0.9\textwidth]{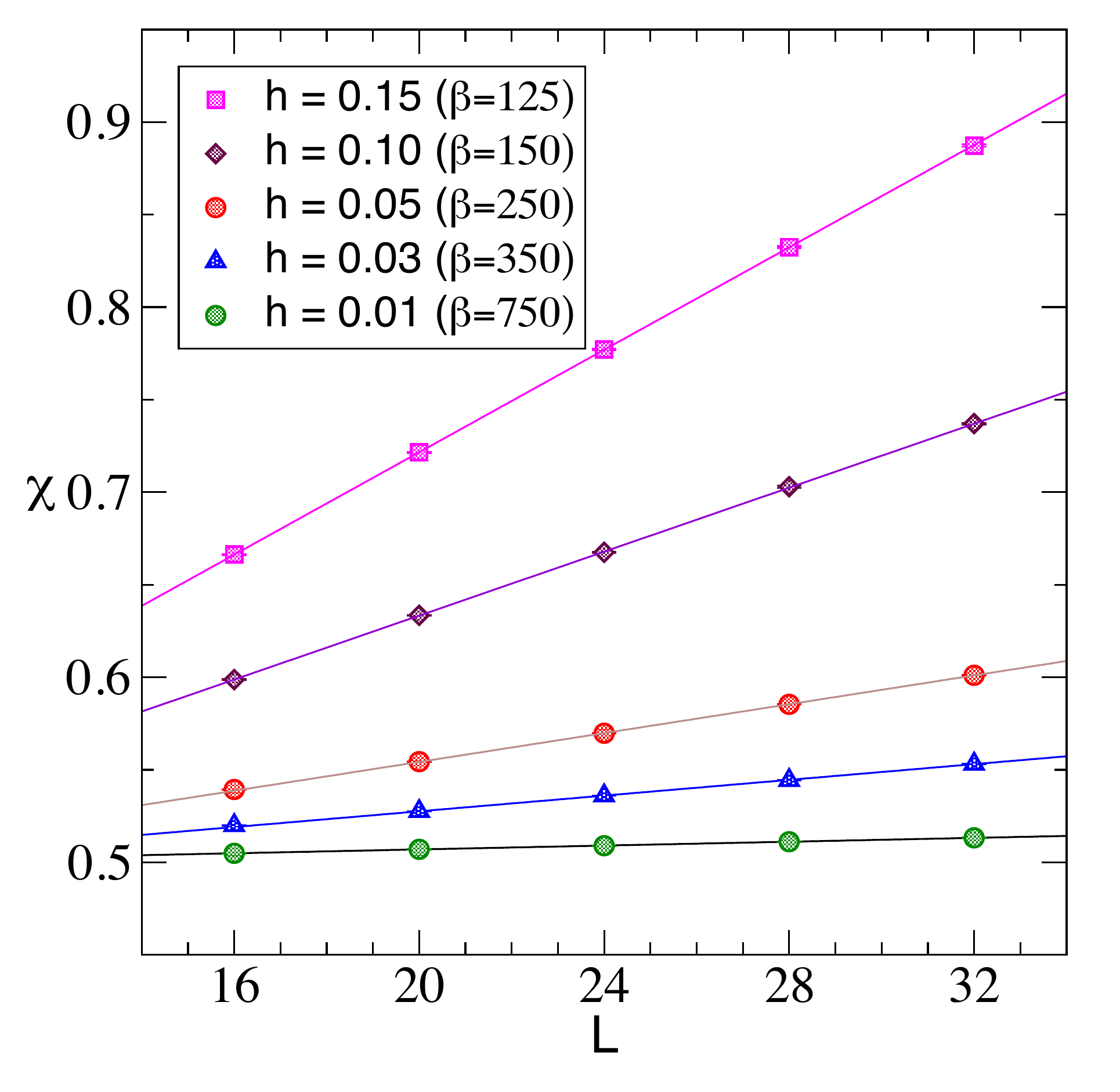}
    \caption{\label{fig:fig6} Plot of the Ising susceptibility $\chi$ as a function of $L$ for various values of $(h,\beta)$. The solid lines are fits to the form $\chi = \phi^2 L + C$ and the values of $\phi$ and $C$ are tabulated in \cref{tab:tab1}. }
  \end{center}
  \end{minipage}
\end{figure}

% Changes to the new version Dec 27 2019.
On the other hand since in our work we do not impose any Gauss' law constraint, chiral symmetry is never explicitly broken and the fermion mass generation at $h \neq 0$ can be viewed as arising due to our choosing one of the Gauss' law sectors spontaneously. Practically, since in $1+1$ dimensions no symmetries can spontaneously break at finite temperatures even in the thermodynamic limit, both Gauss law sectors are sampled equally in our Monte Carlo. However, at zero temperature one of the sectors is chosen spontaneously when $L\rightarrow \infty$.  This spontaneous chiral symmetry breaking is then observed through a non-zero value of $\phi$. In our Monte Carlo calculation, it can be extracted using the chiral susceptibility
\begin{align}
\chi \ = \ \frac{1}{L}\ \Bigg\langle \Big(\sum_j H^{(m)}_j\Big)^2 \Bigg\rangle,
\end{align}
which is expected to scale as $\chi = \phi^2 L + A$ in the symmetry broken phase. In \cref{fig:fig6} we show our data for $\chi$ at various values of $(h,\beta)$. The solid lines are fits to the expected form in the broken phase and the extracted values of $\phi$ and $A$ are tabulated in \cref{tab:tab1}. Thus we see that when $h\neq 0$, our model describes a chirally broken massive fermion phase. Since the energy of the gauge string connecting two fermions excited over the ground state in a given Gauss' law sector will grow with $h$, fermions remain confined. Thus at non-zero couplings our model describes massive confined fermions.

\begin{table}[h]
  \begin{center}
    \begin{small}
\begin{tabular}{c|l|l|l|l|l}
\hline
  ($h,\beta$) & $\phi$ & $A$ & $M_w$ & $B$ & $\langle \sigma^z\rangle$\\
\hline
(0.01,750) & 0.0228(3) & 0.4965(2) & 0.180(2) & 0.58(3) & 0.39(1) \\
(0.03,350) & 0.0462(2) & 0.4851(3) & 0.345(2) & 1.22(3) & 0.464(3) \\
(0.05,250) & 0.0625(2) & 0.4763(4) & 0.470(5) & 1.9(2) & 0.500(2) \\
(0.10,150) & 0.0930(2) & 0.4605(5) & 0.685(14) & 2.4(4) & 0.555(1) \\
(0.15,125) & 0.1176(1) & 0.4449(7) & - & - & 0.592(1) \\
\hline  
\end{tabular}
\end{small}
\end{center}
\caption{\label{tab:tab1} Values of the fit parameters $\phi$, $A$,$M_w$ and $B$ explained in the text for various values of $(h,\beta)$.}
\end{table}

\begin{figure}
  \begin{minipage}[t]{0.45\textwidth}
    \begin{center}
\includegraphics[width=\textwidth]{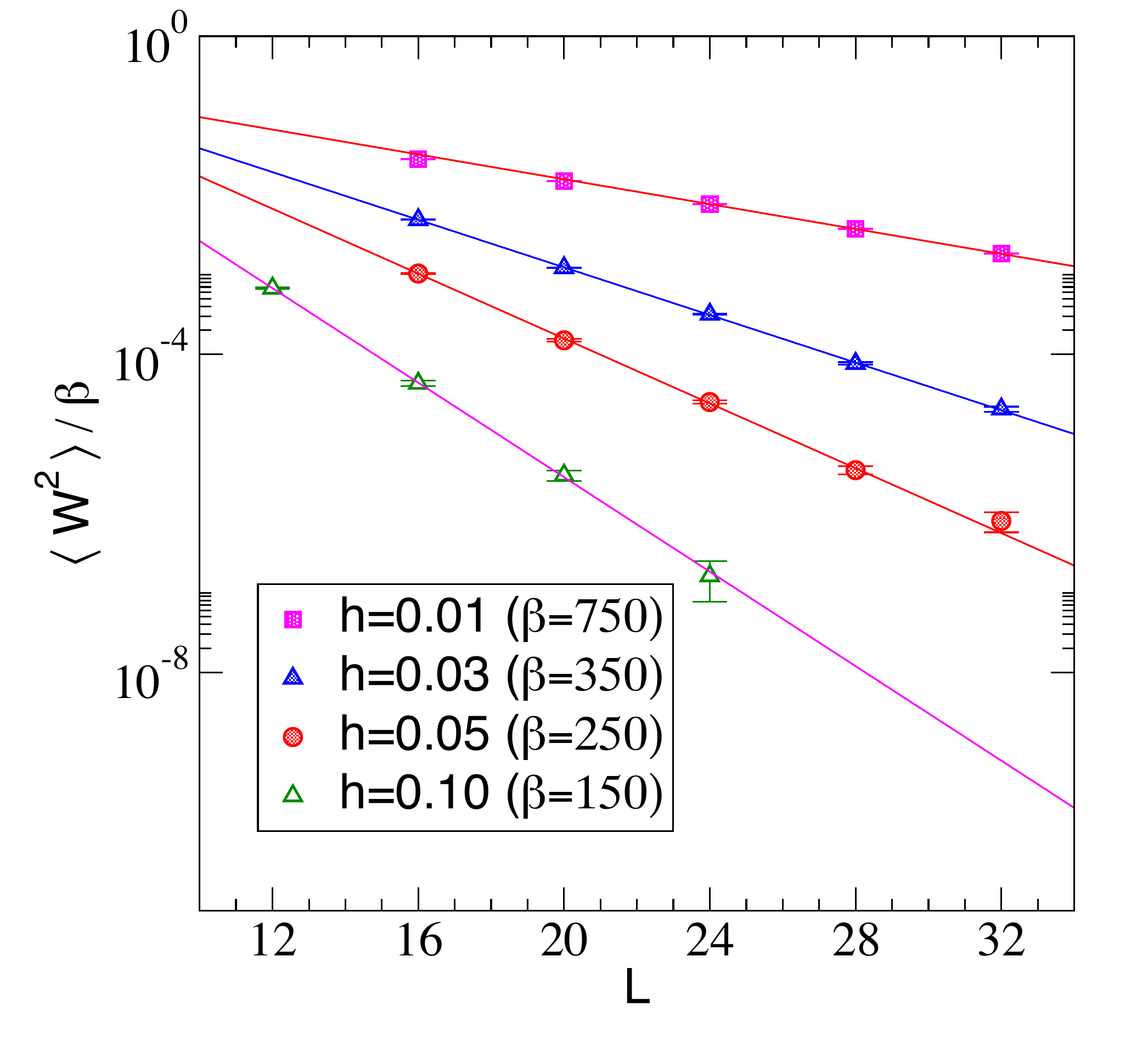}
\caption{\label{fig:fig8} Plot of winding number susceptibility $\langle W^2\rangle$ as a function of $L$ for various values of $(h,\beta)$. The solid lines are fits to the form $\langle W^2\rangle = A \beta \exp(-M_w L) $ and the values of $M_w$ and $A$ are tabulated in \cref{tab:tab1}..}
\end{center}
\end{minipage}
\hspace{1cm}
\begin{minipage}[t]{0.45\textwidth}
  \begin{center}
\includegraphics[width=0.9\textwidth]{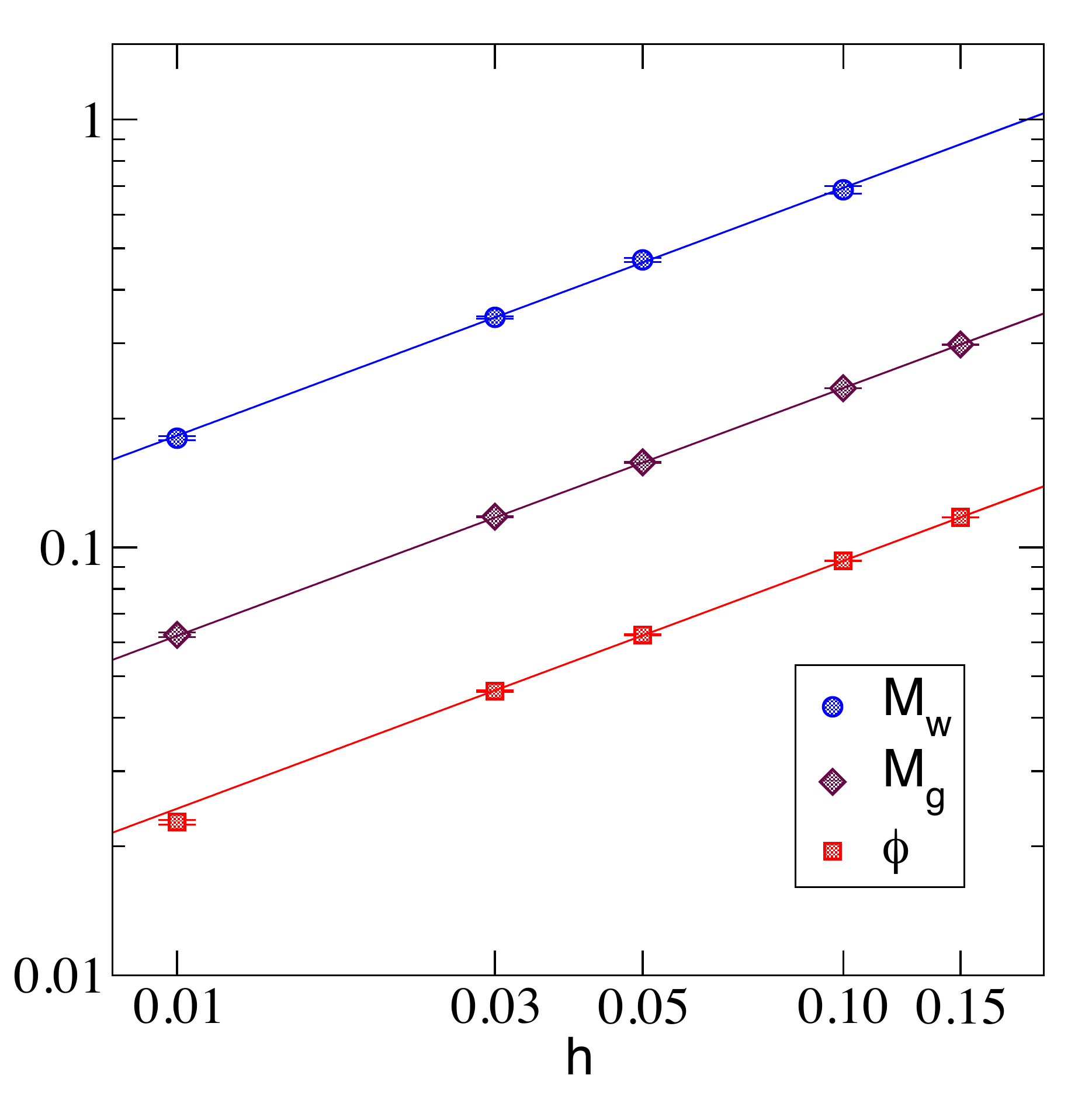}
\caption{\label{fig:fig7} Plot of the winding mass $M_w$, the gauge mass $M_g$ as a function of $h$ from \cref{tab:tab1}. The solid lines are fits to the form $C\ h^{0.58}$, by dropping the $h=0.01$ data in the case of the chiral order parameter for reasons explained in the text.}
\end{center}
\end{minipage}
\end{figure}

\section{Critical Exponent and Scaling}

The deconfined quantum critical point at $h=0$ can be used to define a continuum limit of our massive $Z_2$ lattice gauge theory \cite{Brezin:1981zs}. The effective action of the two dimensional continuum quantum field theory (QFT) that emerges will be described a theory of massless fermions with a relevant coupling at leading order of the form
\begin{align}
  S_h = h \int d^2x \ {\cal O}_h.
\end{align}
As $h$ vanishes we expect the chiral order parameter to vanish as $\phi \sim h^p$, where $p$ is a critical exponent (anomalous dimension) that depends on the  dimensions of $\phi$ and ${\cal O}_h$ at the critical point.

The simplest possibility is that ${\cal O}_h$ is just a fermion bilinear mass term. This is consistent with the fact that chiral symmetry is broken in each fixed Gauss' law sector that emerges when $h\neq 0$. Since the chiral order parameter depends linearly on the fermion mass, this implies that $p=1$. Of course ${\cal O}_h$ will also contain four-fermion couplings, but these would be less relevant and can be ignored in the current discussion. On the other hand, since the lattice Hamiltonian \cref{eq:model1} contains a long range interaction $F_j$, we cannot rule out the possibility that ${\cal O}_h$ contains a term that is more relevant than a mass term. If this is true then $p < 1$ is also possible.

We can study this question by fitting the values of $\phi$ in \cref{tab:tab1} to the form $\phi \sim h^p$. We find an excellent fit for $p = 0.578(2)$ if we drop the $h=0.01$ point, which is justified since $P(Q_e)$ has not yet saturated to $0.5$ at $(h=0.01,\beta=750)$ (see \cref{fig:fig4}). Although it is possible that the range of $h$ we have used in the fit is influencing the value of $p$, the fact that it is considerably smaller than $1$ suggests that the long range part of $F_j$ is indeed playing an interesting role.

At the free massless fermion fixed point in two dimensions, the chiral order parameter $\phi$ has the canonical dimensions of a mass since it is a fermion bilinear. Assuming this argument extends to our deconfined quantum critical point, $\phi \sim h^p$ implies that other quantities with the dimension of mass induced by $h$ will also scale similarly. We have verified this scaling prediction using two other quantities with mass dimensions. We first consider the term $h \langle \sigma^z\rangle$, which is just the average of the Hamiltonian density in one dimension and has dimensions of a square of a mass scale. Thus we can define $M_g = \sqrt{h \langle\sigma^z\rangle}$ which has dimensions of a mass, and according to our above prediction we expect $M_g \sim h^p$. In \cref{tab:tab1} we show our Monte Carlo results for $\langle \sigma^z\rangle$ at various values of $(h,\beta)$. If we use this data to extract $M_g$ and fit it to the form $M_g \sim h^p$ we get $p = 0.577(2)$, in excellent agreement with our result above.

We can also compute a winding mass $M_w$, obtained from the exponential decay of the spatial winding number susceptibilty $\langle W^2\rangle$ with spatial size $L$. Here $W$ is the spatial winding of fermion worldlines of our Monte Carlo configurations. We expect $\langle W^2\rangle/\beta = B \exp(-M_w L) $ at low temperatures. In \cref{fig:fig8} we show our data for $\langle W^2\rangle/\beta$ as a function of $L$ for four different values of $(h,\beta)$. The solid lines are fits of the data to the expected form. The values of $M_w$ and $B$ extracted from these fits are tabulated in \cref{tab:tab1}. Due to an exponentially small signal, we are unable to extract $M_w$ reliably for $h > 0.1$. Fitting the data in the table to $M_w \sim h^p$ we get $p = 0.589(8)$, again in excellent agreement with the previous two results. The data for $\phi$, $M_g$ and $M_w$ as functions of $h$ are shown in \cref{fig:fig7}. We find that $C h^p$, with $p \approx 0.579$ describes all the three mass scales in the region shown.

\section{Conclusions}

In this work we have studied a simple $Z_2$ lattice gauge theory where all local degrees of freedom can be represented by single qubits, so that it can be easily explored on a quantum computer. Our model has an interesting deconfined quantum critical point, which when perturbed by a relevant coupling $h$ leads to a massive quantum field theory with some similarities with QCD. The coupling $h$ seems to be more relevant than a simple fermion mass term. Although we did not impose any Gauss' law in our model, it emerges naturally at low temperatures in the massive phase. Doping the system changes the emergent Gauss' law sectors. Extensions of our work to higher dimensions with bosonic matter fields is easy and would also be interesting especially given the richness of phase diagrams of such theories \cite{PhysRevB.82.085114}. Finally, it would be interesting to study our model using tensor network methods that have been successfully applied to study the Schwinger model recently \cite{PhysRevD.94.085018,PhysRevLett.118.071601,PhysRevD.93.094512}. After our work was published, our model was studied in the $Q_j =1$ Gauss law sector \cite{Borla:2019chl}. The model was solved using bosonization ideas in the large $h$ limit and it was shown that one gets a Luttinger liquid, which seems to extend to all values of $h$. This is quite different from the massive phase we find here in the $Q_j = \pm (-1)^j$ sectors.

\section*{Acknowledgments}

We would like to thank Fakher Assaad for helpful discussions and also for sharing some of his notes on the subject, which was helpful to us in our derivations in \cref{sec2}. We would also like to thank Ashvin Vishwanath, Uwe-Jens Wiese, and Ruben Verresen for helpful comments. The material presented here is based upon work supported by the U.S. Department of Energy (USDOE), Office of Science, Nuclear Physics program under Award Numbers DE-FG02-05ER41368. SC would also like to thank Tanmoy Bhattacharya, Rajan Gupta, Hersh Singh and Rolando Somma for collaboration, which is funded under a Duke subcontract from Department of Energy (DOE) Office of Science - High Energy Physics Contract \#89233218CNA000001 to Los Alamos National Laboratory.

\bibliographystyle{elsarticle-num-names} 
\bibliography{ref,worm,z2}

\clearpage

\appendix

\begin{center}
{\bf Supplementary Material}
\end{center}

\section{Worldline Monte Carlo Algorithm}

Here we briefly explain our Monte Carlo algorithm which is a simple extension of well known algorithms developed earlier to update worldline configurations \cite{Syljuaen:2002zz,Prokofev:2001ddj,Chandrasekharan:2008gp}. In our algorithm we begin with the partition function
\begin{align}
  Z = \mathrm{Tr}(e^{-\beta H_\mu})
\end{align}
where 
\begin{align}
  H_\mu = -\ \sum_j \Big\{ (c^\dagger_j c_{j+1} + c^\dagger_{j+1} c_j)\sigma_j^x\ + h \sigma_j^z  - \mu\ (-1)^{c^\dagger_j c_j} \Big\} \
  \label{eq:modelmu}
\end{align}
is essentially our model with an additional chemical potential term. By tuning the value of $\mu$ we can change the particle number $N_f$ in the ground state. This helps us understand how emergent Gauss' law sectors change due to doping.

In order to express the partition function as a sum over weights of worldline configurations, we first divide $\beta$ into $L_T$ equal slices of width $\varepsilon$ and then express the small imaginary time evolution operator as $e^{-\varepsilon H_\mu} = e^{-\varepsilon H_o}\ e^{-\varepsilon H_e}$, where
\begin{align}
H_e \ =\  - \sum_{j \ \in\ \mbox{\small{even}}} \Big\{ (c^\dagger_j c_{j+1} + c^\dagger_{j+1} c_j)\sigma_j^x\ +  
h \sigma_{j+1}^z - \frac{\mu}{2} \big[(-1)^{c^\dagger_jc_j}  + (-1)^{c^\dagger_{j+1} c_{j+1}}\big] \Big\},
\end{align}
and $H_o$ is a similar operator with the sum over $j$ performed over odd sites. Note that both these operators contain a sum over terms that act on bonds between neighboring sites that commute with each other. Hence one can easily compute the matrix elements of $e^{-\varepsilon H_o}$ and $e^{-\varepsilon H_e}$ as a product of weights on local two dimensional plaquettes as shown in the left figure of \cref{fig:fig1s}. We distinguish two types of plaquettes: fermion plaquettes (shown with darker shade) and gauge plaquettes (shown with lighter shade). The weight of a fermion plaquette is obtained from the transfer matrix elements
\begin{align}
  \Big\langle n_j n_{j+1} s_j |
\exp\Big(\varepsilon\Big\{(c^\dagger_j c_{j+1} + c^\dagger_{j+1} c_j)\sigma_j^x\  - \frac{\mu}{2} \big[(-1)^{c^\dagger_j c_j}  + (-1)^{c^\dagger_{j+1}c_{j+1}}\big] \Big\}\Big) | n'_j, n'_{j+1} s'_j \Big\rangle
\end{align}
where $|n_j,n_{j+1},s_j\rangle$ are eigenstates of $c^\dagger_jc_j$, $c^\dagger_{j+1}c_{j+1}$ and $\sigma^z_j$. Note that these matrix elements have non-zero entries in both diagonal and off-diagonal terms. On the other hand the weight of the gauge plaquette is based on the matrix elements
\begin{align}
  \Big\langle n_j n_{j+1} s_j |
\exp\Big(\varepsilon h \sigma_j^z\Big)  | n'_j, n'_{j+1} s'_j \Big\rangle.
\end{align}
that are always diagonal. For this reason the values $s_j$ across a gauge plaquette are constrained to be the same. As shown in \cref{fig:fig1s} the fermion and gauge plaquettes occur alternately on each time slice.

\begin{figure}[t]
\begin{center}
\includegraphics[width=0.5\textwidth]{fig1.pdf}
\hspace{1cm}
\includegraphics[width=0.4\textwidth]{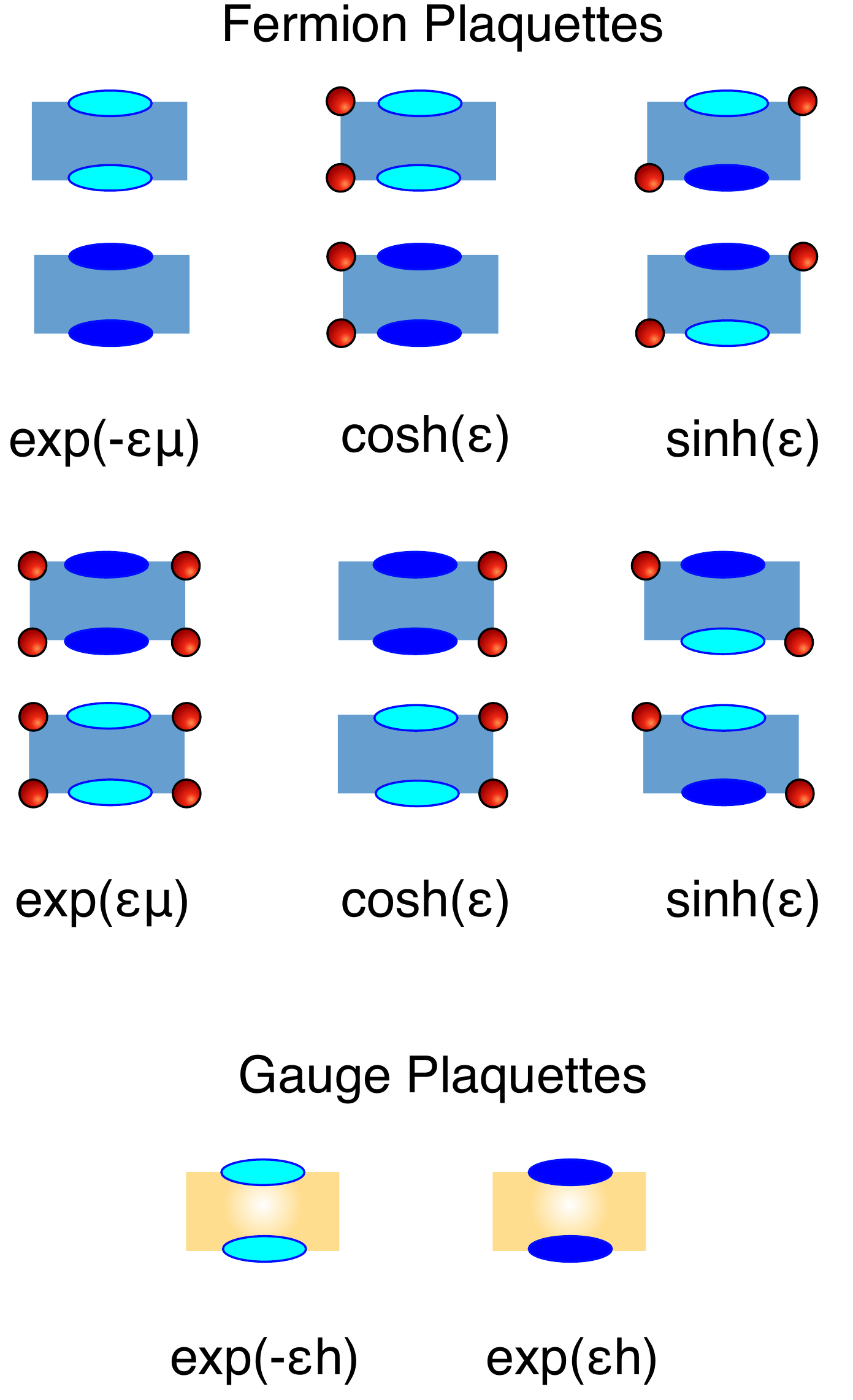}
\caption{\label{fig:fig1c} On the left we give an illustration of the worldline configuration of fermions (filled circles mean $n_j=1$ and absence of circles mean $n_j=0$) and gauge fields (dark $s_j=1$ and light $s_j=-1$ ovals). The weight of a configuration is equal to the product of weights of each plaquette, shown in the right figure.}
\end{center}
\end{figure}

\begin{figure}[t]
\begin{center}
\includegraphics[width=0.45\textwidth]{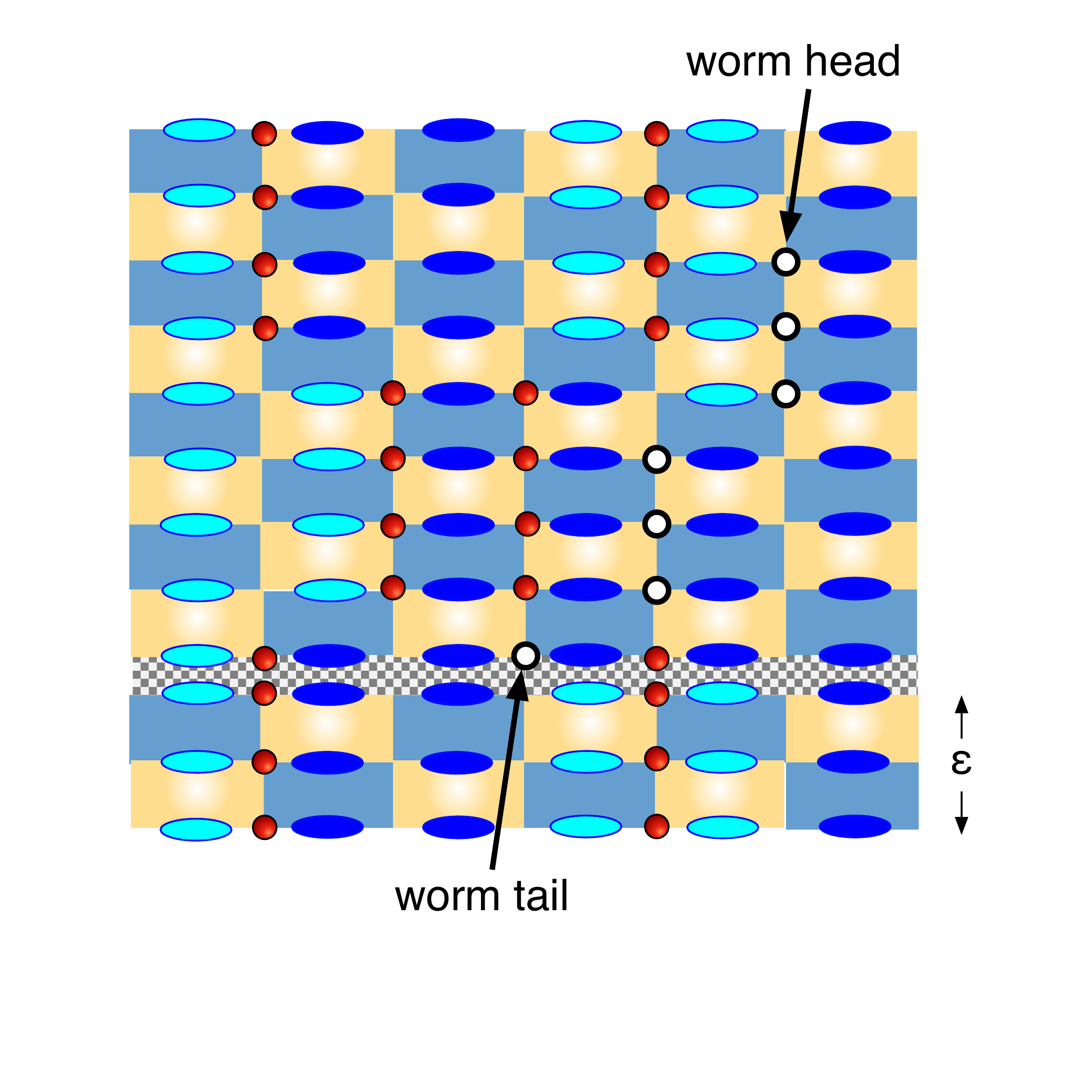}
\caption{\label{fig:fig1s} The figure shows a configuration in the worm sector. Note that in this sector there is an additional temporal boundary (shown by plaquettes shaded differently). The gauge fields across these boundary plaquettes are not constrained to be the same since the trace condition is not imposed on the gauge space in the worm sector, although fermions are still periodic in time. The open circles are fermions introduced during the worm update.}
\end{center}
\end{figure}

During a worm update, we pick a site at random and propose to place a creation operator on that site (worm tail) and place an annihilation operator on one of the other three sites associated with a fermion plaquette (worm head). If this move is accepted according to detailed balance, we accept it as a new configuration. If the configuration is accepted we pick a neighboring fermion plaquette connected to the site containing the annihilation operator (that can be either forward or backward in time) and perform the same operation of adding a creation and annihilation operator. However, now the creation operator is added to the site connected to the previous plaquette and the annihilation operator is chosen to be one of the remaining three sites. If the proposal is accepted the worm head is moved to the new site and the update proceeds. If the proposal is rejected we go back to the previous plaquette and again the update proceeds. Note that these configurations with a worm head and a tail are not part of the partition function since they contain an extra creation and annihilation operator separated by some distance. We refer to these configurations as being in the worm sector. The worm update may only end when the worm head meets the worm tail and we get an acceptable configuration in the partition function sector.

The above steps in constructing the worm update are standard, except that in a gauge theory new complications arise and need to be addressed. The local gauge constraints will be violated at each step unless the gauge links are also simultaneously updated. However, it is impossible to remain in the same Gauss' law sector if only a single fermion is created on a time slice. This is because every fermion creation changes the Gauss' law sector. However, since we allow all Gauss' law sectors to be sampled this is not a problem for our study. On the other hand we do still have to update the gauge links associated to a fermion plaquette whenever the worm head moves to the neighboring site. This is because such a move introduces or elimates a fermion hop and hence will add or remove one $\sigma_j^x$ operator between sites $j$ and $j+1$. But addition of this operator at one time slice has a non-local effect on all time slices in time. Further  the trace constraint of the partition function will be violated.

For this reason in our update we do not impose the trace constraint in the worm sector. We allow the gauge links across a fixed temporal boundary to fluctuate freely.  This temporal boundary is chosen to be on the time slice of the tail site for the worm update. A partial worm update along with the temporal boundary chosen is illustrated on the right in \cref{fig:fig1s}. This then allows us to perform a local worm update in the fermion sector as discussed above, but such an update always includes the possibility of an extra update of all the gauge links associated with $\sigma_j^z$--which is non-local in time--when the fermion plaquette between sites $j$ and $j+1$ is updated. This makes our worm update non-local in time but local in space.

\section{Observables and Tests}

We have measured four observables in our Monte Carlo calculations. These are
\begin{enumerate}
\item Average $Z_2$ field
\begin{align}
  \langle \sigma^z\rangle =\frac{1}{Z L}
\mathrm{Tr}\Big( \Big\{ \sum_j \sigma_j^z \Big\} e^{-\beta H_\mu}\Big)
\end{align}
\item Chiral Susceptibility
\begin{align}
  \chi =\frac{1}{Z L}
\mathrm{Tr}\Big(  \big\{ \sum_j (-1)^{j+n_j} \big\}^2 e^{-\beta H_\mu}\Big)
\end{align}
\item Winding Number Susceptibility
\begin{align}
  \langle W^2\rangle = -\frac{1}{Z}\  \frac{\partial^2}{\partial \theta^2}
\mathrm{Tr}\Big( e^{-\beta H^\theta_\mu}\Big)\Big|_{\theta=0}
\end{align}
where $H_\mu^\theta$ has an extra factor $e^{i\theta}$ or $e^{-i\theta}$ when a fermion hops across the boundary depending on the direction of the hop.
\item Average fermion number
  \begin{align}
  N_f = \frac{1}{Z}\  
\mathrm{Tr}\Big(   \big\{ \sum_j c^\dagger_j c_j \big\}  e^{-\beta H_\mu}\Big)
  \end{align}
\end{enumerate}
We have tested the algorithm by computing these observables exactly on a $L=4$ lattice. In \cref{tab:tab2} we compare our exact answers with those obtained using the algorithm for $\varepsilon=0.01$ (referred to as MC1) and for $\varepsilon = 0.1$ (referred to as MC2). Note that there is not much difference between the two results and both of them match well with the exact calculations. For this reason in our work we have fixed $\varepsilon=0.1$ in our calculations.

\begin{table}
\begin{center}
\begin{tabular}{l|l|l|l|l}
\hline 
\multicolumn{5}{c}{$h=1.0$, $\mu=0.3$, $\beta=5$} \\
\hline
& $\langle \sigma^z\rangle$ & $ N_f $ & $\chi$ & $\langle W^2\rangle $ \\
\hline
Exact & 0.862396...& 2.725711... &0.354376...& 0.001936 \\
MC1 & 0.86222(9) & 2.7251(5) & 0.35461(18) & 0.00179(6) \\
MC2 & 0.86241(9) & 2.7263(5) & 0.35409(17) & 0.00191(6) \\
\hline
& $P_0$ & $P_1$ & $P_2$ & $P_3$ \\ 
\hline
Exact &$0.000011...$ & $0.000398...$ & $0.000398...$ & 0.017363...\\
MC1 & 0.000012(1) & 0.000406(7) & 0.000405(7)& 0.01730(6) \\
MC2 & 0.000012(1) & 0.000398(7) & 0.000403(6)& 0.01736(6) \\
\hline
& $P_4$ & $P_5$ & $P_6$ & $P_7$ \\ 
\hline
Exact & $0.000398...$ &0.13162...&0.01736...&0.15107...\\
MC1 & 0.000393(6) & 0.1318(2)& 0.01733(7) & 0.1512(3)\\
MC2 & 0.000390(7) & 0.1313(2)& 0.01742(8) & 0.1512(3)\\
\hline
& $P_8$ & $P_9$ & $P_{10}$ & $P_{11}$ \\ 
\hline
Exact & $0.000398...$ &0.01736...&0.13162...& 0.15107...\\
MC1 & 0.000396(6) & 0.01737(7) & 0.1317(2) & 0.1507(4) \\
MC2 & 0.000404(7) & 0.01732(6) & 0.1313(2) & 0.1514(2) \\
\hline
& $P_{12}$ & $P_{13}$ & $P_{14}$ & $P_{15}$ \\ 
\hline
Exact & 0.01736...&0.15107...&0.15107...& 0.06141...\\
MC1 &0.01732(7) & 0.1515(4) & 0.1512(4) & 0.0610(2) \\
MC2 &0.01737(7) & 0.1513(3) & 0.1509(3) & 0.0614(2) \\
\hline 
\end{tabular}
\caption{\label{tab:tab2} Comparison between exact calculations and those obtained using our Monte Carlo methods, MC1 (with $\varepsilon=0.01$) and MC2 (with $\varepsilon=0.1$). The definition of observables shown are discussed in the text. The lattice size is $L_x=4$.}
\end{center}
\end{table}

\begin{figure}
\begin{minipage}[t]{0.45\textwidth}
\begin{center}
\includegraphics[width=0.87\textwidth]{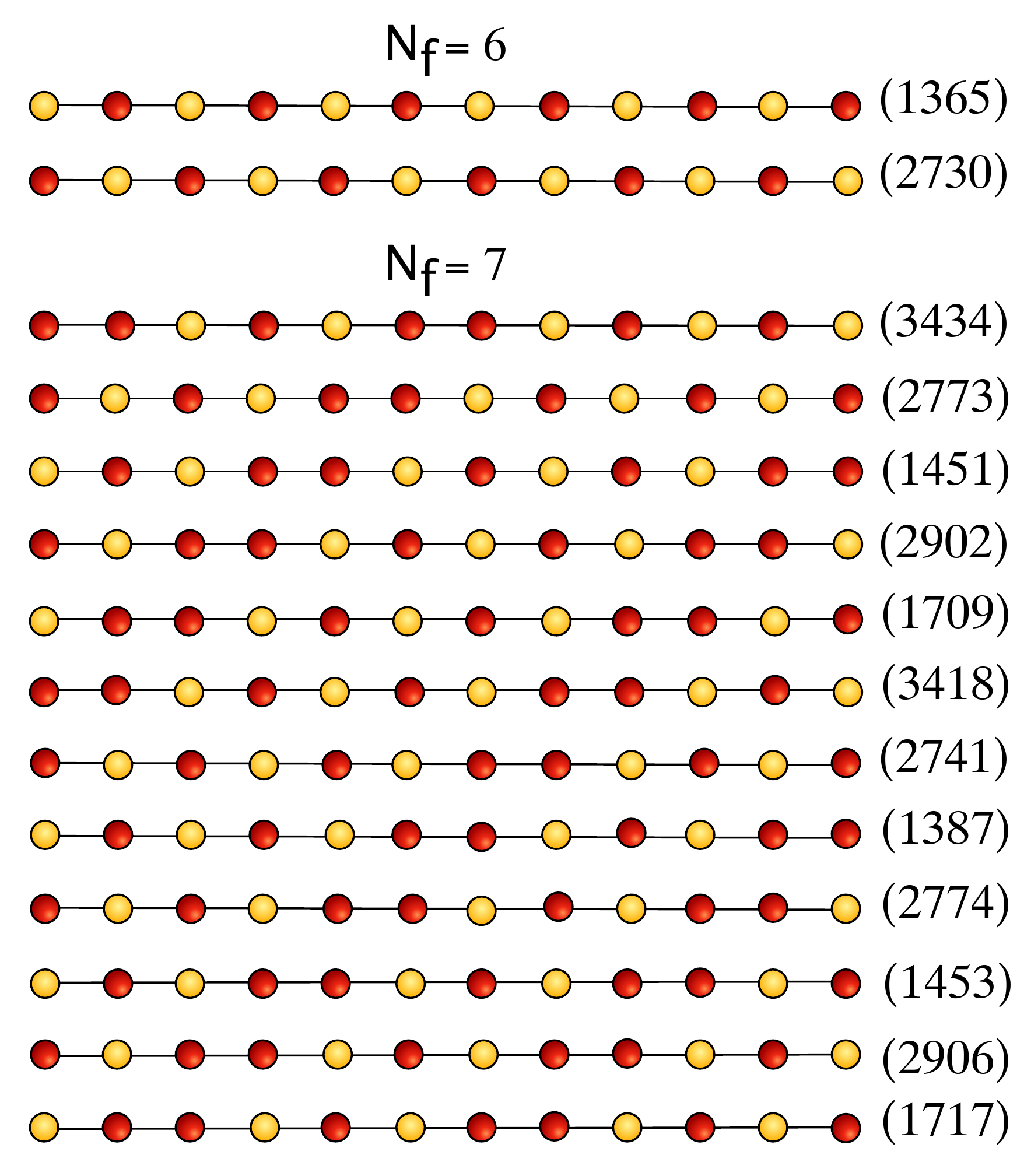}
\caption{\label{fig:fig9} The distribution of $Q_j$'s in the emergent Gauss' law sectors when $N_f=6$ and $N_f=7$ on a $L=12$ lattice when $h=0.05$. The darker and lighter circles stand for $Q_j=1$ and $Q_j=-1$ respectively.}
\end{center}
\end{minipage}
\hspace{1cm}
\begin{minipage}[t]{0.45\textwidth}
  \begin{center}
\includegraphics[width=\textwidth]{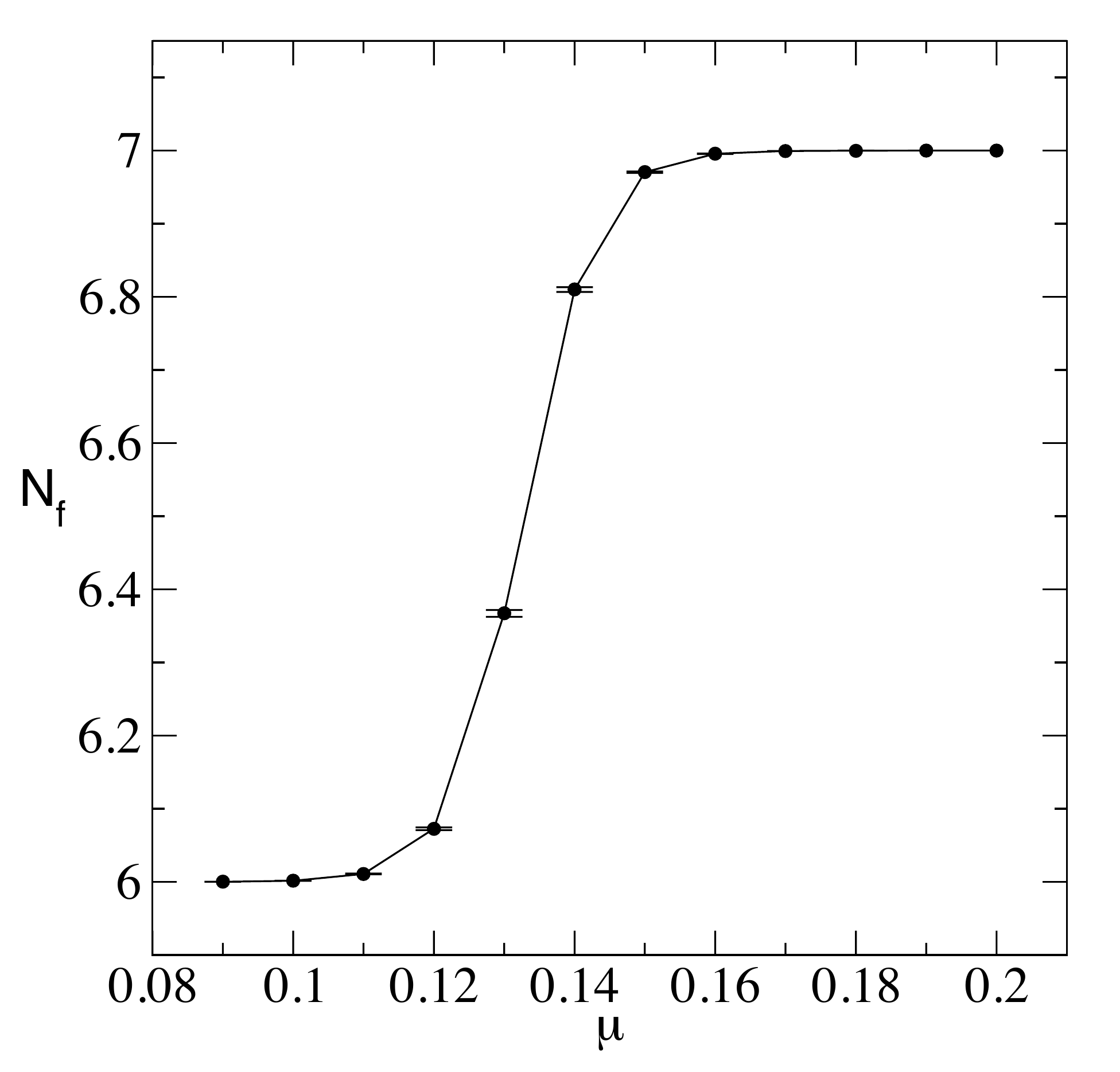}
\caption{\label{fig:fig10} Plot of the average number of fermions $N_f$ as a function of the chemical potential $\mu$ at $h=0.05,\beta = 100$ and $L=12$.}
\end{center}
\end{minipage}
\end{figure}

\begin{figure}
\begin{minipage}[t]{0.45\textwidth}
\begin{center}
\includegraphics[width=0.93\textwidth]{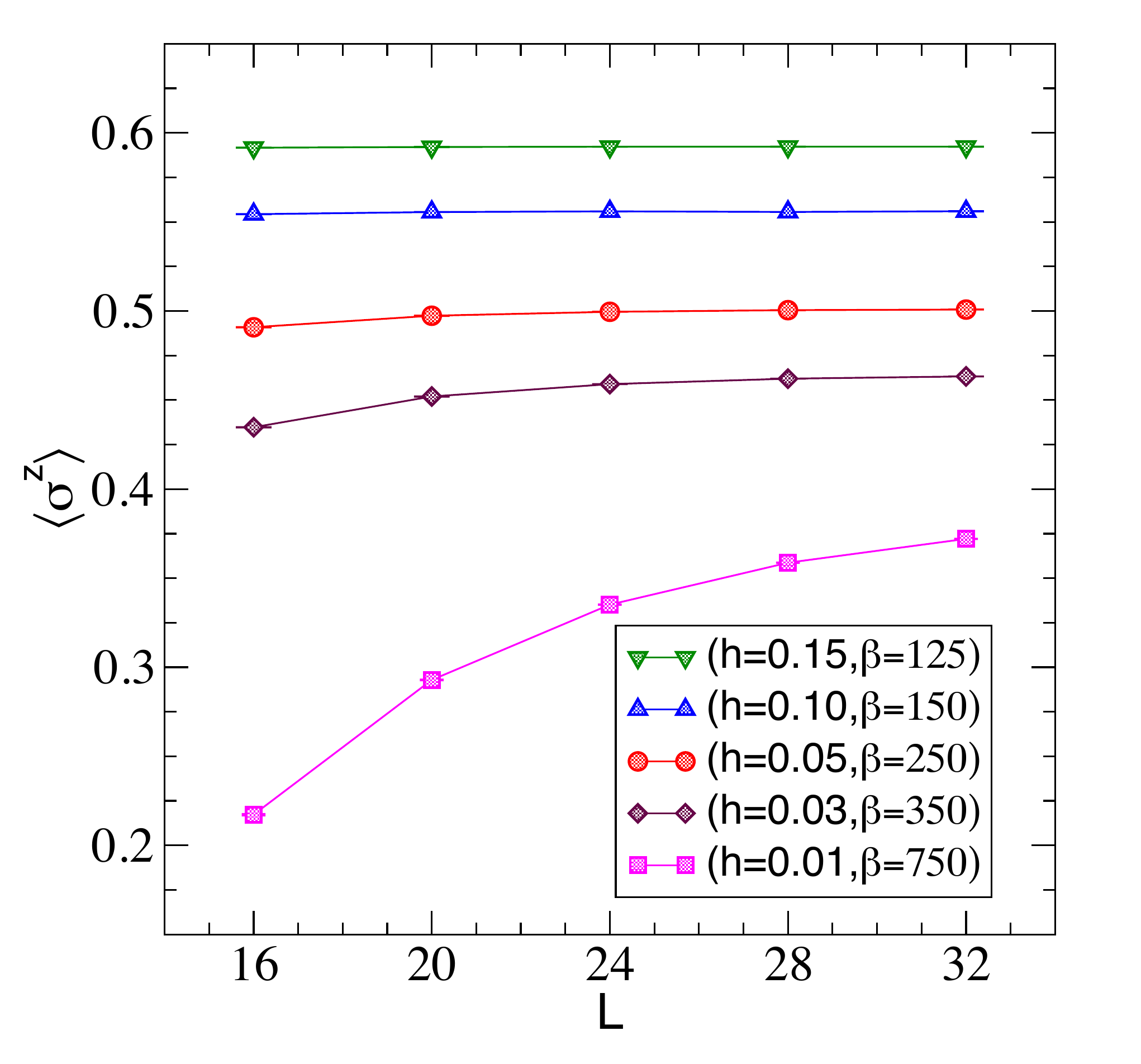}
\caption{\label{fig:fig11} Plot of $\langle \sigma^z\rangle$ as a function of $L$ at various values of $(h,\beta)$. We take the $L=32$ data to estimate the numbers used in the main paper.}
\end{center}
\end{minipage}
\hspace{1cm}
\begin{minipage}[t]{0.45\textwidth}
  \begin{center}
\includegraphics[width=\textwidth]{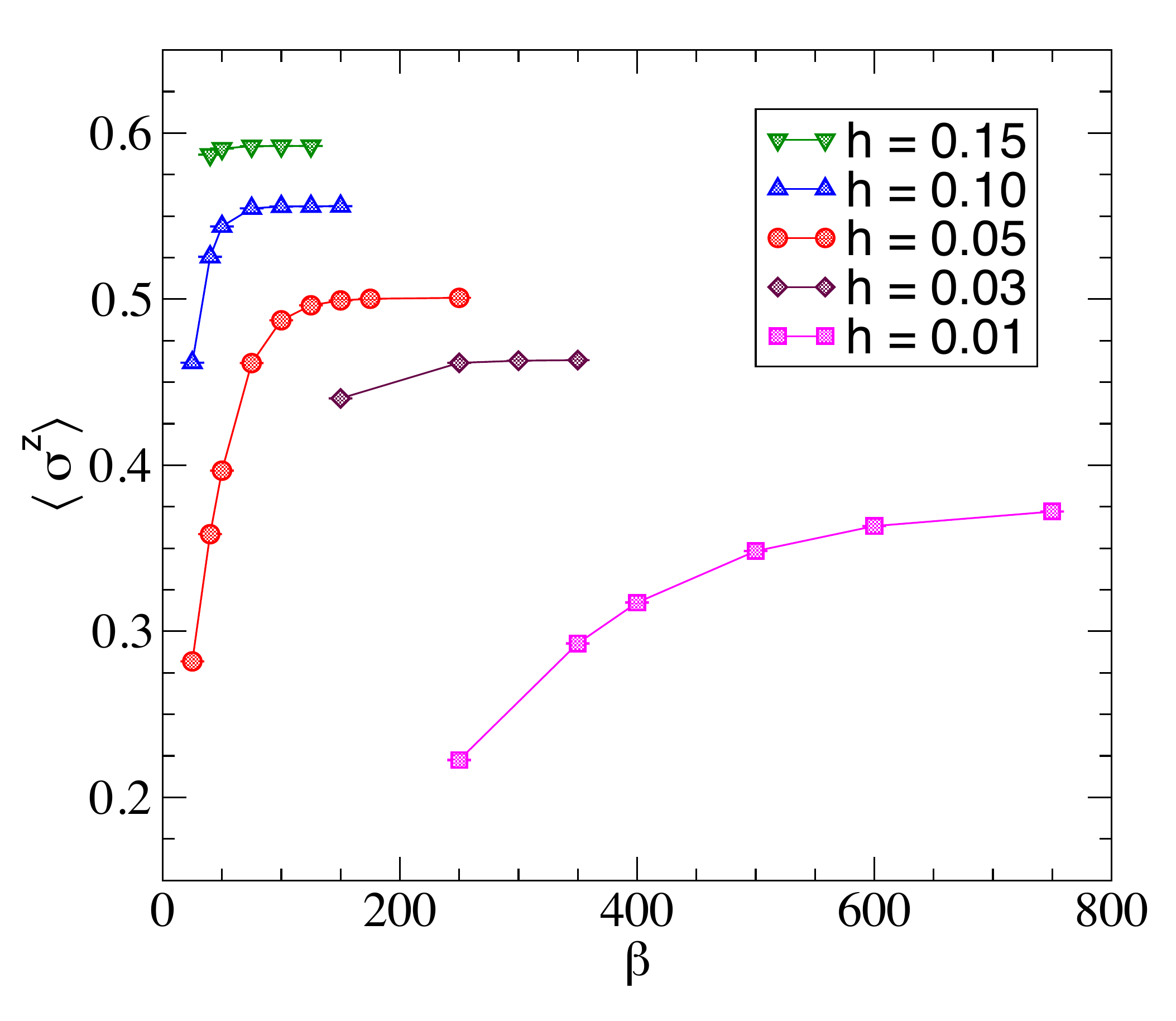}
\caption{\label{fig:fig12} Plot of $\langle \sigma^z\rangle$ as a function of $\beta$ at $L=32$. This shows that except for $h=0.01$, our data has saturated to the zero temperature values at the largest $\beta$ values for each $h$.}
\end{center}
\end{minipage}
\end{figure}

\section{Summary of Additional Plots}

In order to document some of our results in a more complete form, we provide several clarifying figures.
\begin{enumerate}
\item In \cref{fig:fig9} we give a pictorial illustration of the Gauss' law sectors that emerge with $6$ and $7$ fermions on an $L=12$ lattice.
\item In \cref{fig:fig10} we plot the average fermion number as a function of the chemical potential on an $L=12$ lattice when $h=0.05$ and $\beta=100$. In the main paper we illustrate the physics at $\mu=0.2$ where $N_f=7$.
\item In \cref{fig:fig11} and \cref{fig:fig12} we show the dependence of $\langle \sigma^z\rangle$ as a function of $L$ and $\beta$ for various values of $h$. The values of $\langle \sigma^z\rangle$ we use in the main paper are estimated from this data.
\end{enumerate}

\end{document}